\documentclass[12pt]{article}

\usepackage{scicite}
\usepackage{times}
\usepackage{graphics}

\newenvironment{sciabstract}{%
\begin{quote} \bf}
{\end{quote}}

\newcounter{lastnote}
\newenvironment{scilastnote}{%
\setcounter{lastnote}{\value{enumiv}}%
\addtocounter{lastnote}{+1}%
\begin{list}%
{\arabic{lastnote}.}
{\setlength{\leftmargin}{.22in}}
{\setlength{\labelsep}{.5em}}}
{\end{list}}

\title{Controlled Single-Photon Emission from a Single Trapped
Two-Level Atom}

\author
{B. Darqui\'e, M.P.A. Jones, J. Dingjan, J. Beugnon, S. Bergamini, \\
Y. Sortais, G. Messin, A. Browaeys$^{\ast}$, P. Grangier \\
\\
\normalsize{Laboratoire Charles Fabry de l'Institut d'Optique (UMR 8501),} \\
\normalsize{Bat. 503, Centre Universitaire, 91403 Orsay, France}\\
\\
\normalsize{$^{\ast}$To whom correspondence should be addressed; E-mail:
antoine.browaeys@iota.u-psud.fr}
}

\date{}

\begin{document}

\baselineskip24pt

\maketitle

\begin{sciabstract}
By illuminating an individual rubidium atom stored in a tight
optical tweezer with short resonant light pulses, we create an
efficient triggered source of single photons with a well-defined
polarization. The  measured intensity correlation of the emitted
light pulses exhibits almost perfect antibunching. Such a source
of high rate, fully controlled single photon pulses has many
potential applications for quantum information processing.
\end{sciabstract}


\clearpage

Implementing a deterministic or conditional two qubit quantum gate
is a key step towards quantum computation. Deterministic gates
generally require a strong interaction between the particles that
are used to carry the physical qubits~\cite{Zoller}. Recently,
controlled-not gates have been realized using trapped ions and
incorporated in elaborate quantum algorithms
\cite{teleportation_wineland,teleportation_blatt,
code_correcteur_wineland}. So far, individually addressed two
qubit gates have not been demonstrated with neutral atoms.
Promising results have been obtained on entangling neutral atoms
using cold controlled collisions in an optical lattice
\cite{bloch}, but the single qubit operations are difficult to
perform in such a system.

Another approach is to bypass the requirement for a direct
interaction between the qubits, and use instead an interference
effect and a measurement-induced state projection to create the
desired operation \cite{KLM,Dowling}. An interesting recent
development of this idea is to use photon detection events for
creating entangled states of two atoms
\cite{protsenko02b,simon,duan03}. This provides ``conditional"
quantum gates, where the success of the logical operation is
heralded by appropriate detection events. These schemes can be
extended to realize a full controlled-not gate, or a Bell-state
measurement, or more generally to implement conditional unitary
operations \cite{protsenko02b,Beige}. They could be implemented by
using, for instance, trapped ions \cite{Blinov}, or atoms in
microscopic dipole traps. These proposals require the controlled
emission of indistinguishable single photons by at least two
identical emitters.

Various single photon sources have been implemented using
solid-state systems as well as atoms or ions. Solid-state systems
such as single molecules, nitrogen-vacancy centers in diamond or
quantum dots allow high single photon rates
\cite{source_photon_unique}. However, realizing truly identical
sources is a major problem for such systems, due to
inhomogeneities in both the environment of the emitters, and the
emitters themselves. Another approach is provided by sources based
on neutral atoms \cite{rempe2002,kimble2004} or ions
\cite{walther2004} strongly coupled to a mode of a high-finesse
optical cavity. Such sources are spectrally narrow, and the
photons are emitted into a well-defined spatial mode, thus opening
the way to coherent coupling of the quantum state of single atoms
and single photons. However, the rate at which the system can emit
photons is limited by the cavity and is often low in practice.
Moreover, the need to achieve the strong coupling regime of cavity
quantum electrodynamics remains a demanding experimental
requirement.

We present a triggered single-photon source
based on a single rubidium atom trapped at the focal point of a
high-numerical-aperture lens (N.A. = $0.7$). We also
show that we have full control of the optical
transition by observing Rabi oscillations. Under these conditions
our system is equivalent to the textbook model formed by a
two-level atom driven by monochromatic light pulses. Previous
work has shown that by using holographic techniques one can
create arrays of dipole traps, each containing a single atom,
which can be addressed individually \cite{bergamini}. The work
presented here can therefore be directly scaled to two or more
identical emitters.


We trap the single rubidium 87 atom at the focus of the lens using
a far-detuned optical dipole trap (810 nm), loaded from an optical
molasses. The same lens is also used to collect the fluorescence
emitted by the atom (Fig.~1). The experimental apparatus is
described in more detail in references
\cite{Schlosser01,Schlosser02}. A crucial feature of our
experiment is the existence of a ``collisional blockade"
mechanism~\cite{Schlosser02} which allows only one atom at a time
to be stored in the trap: if a second atom enters the trap, both
are immediately ejected. In this regime the atom statistics are
sub-Poissonian and the trap contains either one or zero (and never
two) atoms, with an average atom number of $0.5$.

The trapped atom is excited with $4$ ns pulses of laser light,
resonant with the $S_{1/2},\, F=2 \rightarrow P_{3/2},\,
F^{\prime} = 3$ transition at $780.2$ nm. The laser pulses are
generated by frequency doubling pulses at 1560 nm, generated by
using an electro-optic modulator to chop the output of a
continuous-wave diode laser. A fiber amplifier is used to boost
the peak power of the pulses prior to the doubling crystal. The
repetition rate of the source is $5$ MHz.

Fluorescence photons are produced by spontaneous emission from the
upper state, which has a lifetime of $26$ ns. The pulsed laser
beam is $\sigma^+$-polarized with respect to  the quantization
axis defined by a magnetic field applied during the excitation.
The trapped atom is optically pumped into the $F=2,\, m_F =+2$
ground state by the first few laser pulses. It then cycles on the
$F=2,\, m_F=+2 \rightarrow F^{\prime}=3,\, m_F^{\prime}=+3$
transition, which forms a closed two-level system emitting
$\sigma^+$-polarized photons. Impurities in the polarization of
the pulsed laser beam with respect to the quantization axis,
together with the large bandwidth of the exciting pulse ($250$
MHz), result in off-resonant excitation to the $F'=2$ upper state,
leading to possible de-excitation to the $F=1$ ground state. To
counteract this, we add a repumping laser resonant with the $F=1
\rightarrow F' = 2$ transition. We check that our two-level
description is still valid in the presence of the repumper by
analyzing the polarization of the emitted single photons (see
supporting online text for further details).

The overall detection and collection efficiency for the light
emitted from the atom is measured to be $0.60 \pm 0.04\%$. This is
obtained by measuring the fluorescence rate of the atom for the
same atomic transition driven by a continuous-wave probe beam, and
confirmed by a direct measurement of the transmission of our
detection system (see supporting online text).


For a two-level atom and exactly resonant square light pulses of fixed
duration $T$, the probability for an atom in the ground state to
be transferred to the excited state is $\sin^2(\Omega T/2)$, the
Rabi frequency $\Omega$ being proportional to the square root of
the power. Therefore the excited state population and hence the
fluorescence rate oscillates as the intensity is increased. To
observe these Rabi oscillations, we illuminate the trapped atom
with the laser pulses during $1$ ms. We keep the length of each laser
pulse fixed at $4$ ns, with a repetition rate of $5$ MHz, and
measure the total fluorescence rate as a function of the laser
power. The Rabi oscillations are clearly visible on our results
(see Fig.~2). From the height of the first peak and the calibrated
detection efficiency measured previously, we derive a maximum
excitation efficiency per pulse of $95\pm 5\%$.

The reduction in the contrast of the oscillations at high laser
power is mostly due to fluctuations of the pulsed laser peak
power. This is shown by the theoretical curve in Fig.~2,
based on a simple two-level model. This model shows
that the $10\%$ relative intensity fluctuations that we measured on
the laser beam  are enough to smear out the oscillations as
observed.

The behavior of the atom in the time domain can be studied by
using time resolved photon counting techniques to record the
arrival times of the detected photons following the excitation
pulses, thus constructing a time spectrum. By adjusting the laser
pulse intensity, we observe an adjustable number of Rabi
oscillations during the duration of the pulse, followed by the
free decay of the atom once the laser has been turned off. The
effect of pulses close to $\pi$,  $2\pi$ and $3\pi$ are displayed
as inserts on Fig.~2, and show  the quality of the coherent
control achieved on a single atom.


In order to  use this system as a single photon source, the laser
power is set to  realize a $\pi$ pulse. To maximize the number of
single photons emitted before the atom is heated out of the trap,
we use the following sequence. First, the presence of an atom in
the dipole trap is detected in real-time using its fluorescence
from the molasses light. Then, the magnetic field is switched on
and we trigger an experimental sequence that alternates
$115\,\mu$s periods of pulsed excitation with $885\,\mu$s periods
of cooling by the molasses light (Fig.~3). The repumping laser
remains on throughout, and the trap lifetime during the sequence
is measured to be $34$~ms. After $100$ excitation/cooling cycles,
the magnetic field is switched off and the  molasses is turned
back on, until a new atom is recaptured and the process begins
again. On average, three atoms are captured per second under these
conditions. The average count rate during the excitation is
$9600$~s$^{-1}$, with a peak rate of $29000$~s$^{-1}$
(corresponding to twice the first peak in Fig.~3)

To characterize the statistics of the emitted light, we measure
the second order temporal correlation function, using a Hanbury
Brown and Twiss type set-up. This is done using the beam splitter
in the imaging system (Fig.~1), which sends the fluorescence light
to two photon-counting avalanche photodiodes that are connected to
a high-resolution time-to-digital conversion counting card in a
start-stop configuration (resolution of about 1 ns). The card is
gated so that only photons scattered during the $115\,\mu$s periods
of pulsed excitation are counted, and the number of coincidence
events is measured as a function of delay. The histogram obtained
after $4$ hours of continuous operation is displayed in Fig.~4, and
shows a series of spikes separated by the period of the excitation
pulses ($200$ ns). The $1/e$ half width of the peaks is $27 \pm 3$
ns, in agreement with the lifetime of the upper state. No
background correction is done on the displayed data. The small
flat background is attributed to coincidences between a
fluorescence photon, and an event coming either from stray laser
light (about $175$ counts/sec), or dark counts of the avalanche
photodiodes (about $150$ counts/sec). When these events are
corrected for, the integrated residual area around zero delay is
$3.4\%\, \pm\, 1.2\%$ of the area of the other peaks.

We calculate~\cite{calculs} that under our experimental
conditions, the probability to emit exactly one photon per pulse
is $0.981$ whereas the probability to emit two photons is $0.019$.
These two-photon events would show up in the correlation curve as
coincidences close to zero delay (still with no coincidences at
exactly zero delay). From our calculation, the value for the ratio
of the area around zero delay compared to the others is $3.7\%$,
in excellent agreement with the experimental results.

Finally, we discuss the coherence properties of the emitted
photons, necessary for entanglement protocols based on the
interference between two emitted photons, either from the same
atom or from different atoms. As our collection optics are
diffraction-limited, the outgoing photons should be in a single
spatial mode of the electromagnetic field. As far as temporal
coherence is concerned, the main limiting factor appears to be the
motion of the atom in the trap, which can be controlled by
optimized cooling sequences. We then anticipate that our source
should be Fourier-limited by the lifetime of the excited state. We
are now working to characterize the coherence of our single-photon
source, and to use it to observe multiple atom interference
effects.

\clearpage

\bibliographystyle{Science}

\begin{thebibliography}{30}
\bibitem{calculs} We have performed a full calculation of the
second order correlation function using the Heisenberg-Langevin
equations, which gives the areas of the peaks. We have also
calculated the photon emission probabilities using density matrix
and Monte Carlo methods, which lead to the same result.
\end{thebibliography}

\begin{scilastnote}
\item  We thank Patrick Georges for his assistance in designing
the pulsed laser system. This work was supported by the European
Union through the IST/FET/QIPC project ``QGATES" and the Research
Training Network ``CONQUEST".
\end{scilastnote}

\newpage
\section*{ Supporting Online Material}
www.sciencemag.org\\
Supporting online text

\subsection*{Polarization analysis of the emitted single photons}

As explained in the main text, polarization imperfections lead
to a depumping process to the $F=1$ ground state. A repumping laser is
used to counteract this process and minimize deviations from the
two-level behavior.

\medskip\noindent To check the validity of our two-level description,
we have investigated the effect of impurities in the polarization
of the pulsed laser beam with respect to the quantization axis. We
measure that on  average the atom is pumped into the $F=1$ ground
state by spontaneous emission after 120 excitations. In the
presence of the repumping light, a rate equation model of the
repumping process (including the repumping laser as well as the
pulsed excitation) shows that the atom spends more than 90\% of
its time on the cycling $F=2,m_F=+2 \rightarrow F'=3,m_F'=+3$
transition, as desired.

\medskip\noindent In addition, we have measured the polarization of the
emitted light, using a polarizer to select fluorescence light
either polarized perpendicularly ($\bot$) or parallel ($\|$) to
the quantization axis. For a narrow collection angle, $\bot$ would
correspond to the circularly polarized photons emitted on the
cycling transition, and $\|$ to $\pi$ polarized photons. Here, the
measured contrast $(R_{\bot} - R_{\|})/(R_{\bot}+R_{\|})= 72 (\pm
2)\%$, where $R_{\bot}$ and $R_{\|}$ are the count rates
perpendicular and parallel to the quantization axis respectively.
The largest part of this depolarization is actually due to the
very large numerical aperture (N.A. = 0.7) of the collection lens,
which decreases to $77\,\%$ the maximum contrast obtainable for
purely $\sigma$-polarized fluorescence. From the measured contrast
of $72\%$, we calculate that $3\%$ of the collected photons are
$\pi$-polarized and we attribute this to photons induced by the
depumping-repumping processes. This number is compatible with the
results of the rate equation model discussed in the previous
paragraph.

\subsection*{Collection and detection efficiency}

The overall collection and detection efficiency of $(0.60\pm0.04)\%$ is
obtained by measuring the fluorescence rate of the atom as a function of
the power of a continuous-wave probe beam. Since the saturated photon
emission rate for a closed two-level system is $\Gamma/2$, where
$\Gamma$ is the inverse of the natural lifetime, the collection and
detection efficiency can be obtained directly from the measured count
rate.

\medskip\noindent This value is compatible with that obtained from a
direct evaluation of the transmission of our detection system. The
transmission of our lens is measured to be $87\%$ and its
collection solid angle is $0.15\times 4\pi$~sr. Because the
emission pattern for $\sigma^+$-polarized photons is not
isotropic, the effective solid angle of collection must be
corrected by a factor of $85\%$. The transmission of the optical
elements in the imaging system is $58\%$. Finally the light passes
through a pinhole before illuminating the avalanche photodiode.
The largest uncertainty is in the combination of the pinhole
transmission and photodiode quantum efficiency, which is estimated
to be around 10\%. Multiplying all factors gives an overall
collection and detection efficiency compatible with the $0.6 \%$
quoted above.

\clearpage

\noindent {\bf Figure 1.} Schematic of the experiment. The same lens
is used to focus the dipole trap and collect the fluorescence
light. The fluorescence is separated by a dichroic mirror and
imaged onto two photon counting avalanche photodiodes (APD),
placed after a beam-splitter (BS). The insert shows the relevant
hyperfine levels and Zeeman sublevels of rubidium $87$. The cycling
transition is shown by the arrow. Also shown is the nearby
$F'=2$ level responsible for the depumping.

\

\noindent {\bf Figure 2.} Total count rate (squares) as a function
of the average power of the pulsed beam, for a fixed pulse length
of $4$ ns and a repetition rate of $5$ MHz. The solid line is a
theoretical curve using a simple two-level model that includes
spontaneous emission and intensity fluctuations. The inserts show
the time spectra for the laser intensities corresponding to $\pi$,
$2\pi$ and $3\pi$ pulses.

\

\noindent {\bf Figure 3.} Fluorescence signal measured by one of
the two photodiodes during the experimental sequence, averaged
over $22958$ cycles. Peaks are observed corresponding to the
$115\,\mu$s periods of pulsed excitation, separated by  periods of
lower fluorescence induced by the molasses light during the
$885\,\mu$s of cooling. The exponential decay of the signal is due
the lifetime of the atom in the trap, which is $34$ ms under these
conditions. Insert: A close-up of the signal clearly shows the
alternating excitation and cooling periods.

\

\noindent {\bf Figure 4.} Histogram of the time delays in the
start-stop experiment. The histogram has been binned $4$ times
leading to a $4.7$~ns time resolution. No correction for background
has been made. The absence of a peak at zero delay shows that the
source is emitting single photons. During the $4$-hour experimental
run, $43895$ sequences were completed, which corresponds to a total
of $505$ seconds of excitation.  A total of $4.83\times 10^6$
photons were detected by the two photodiodes.

\begin{figure}
\includegraphics{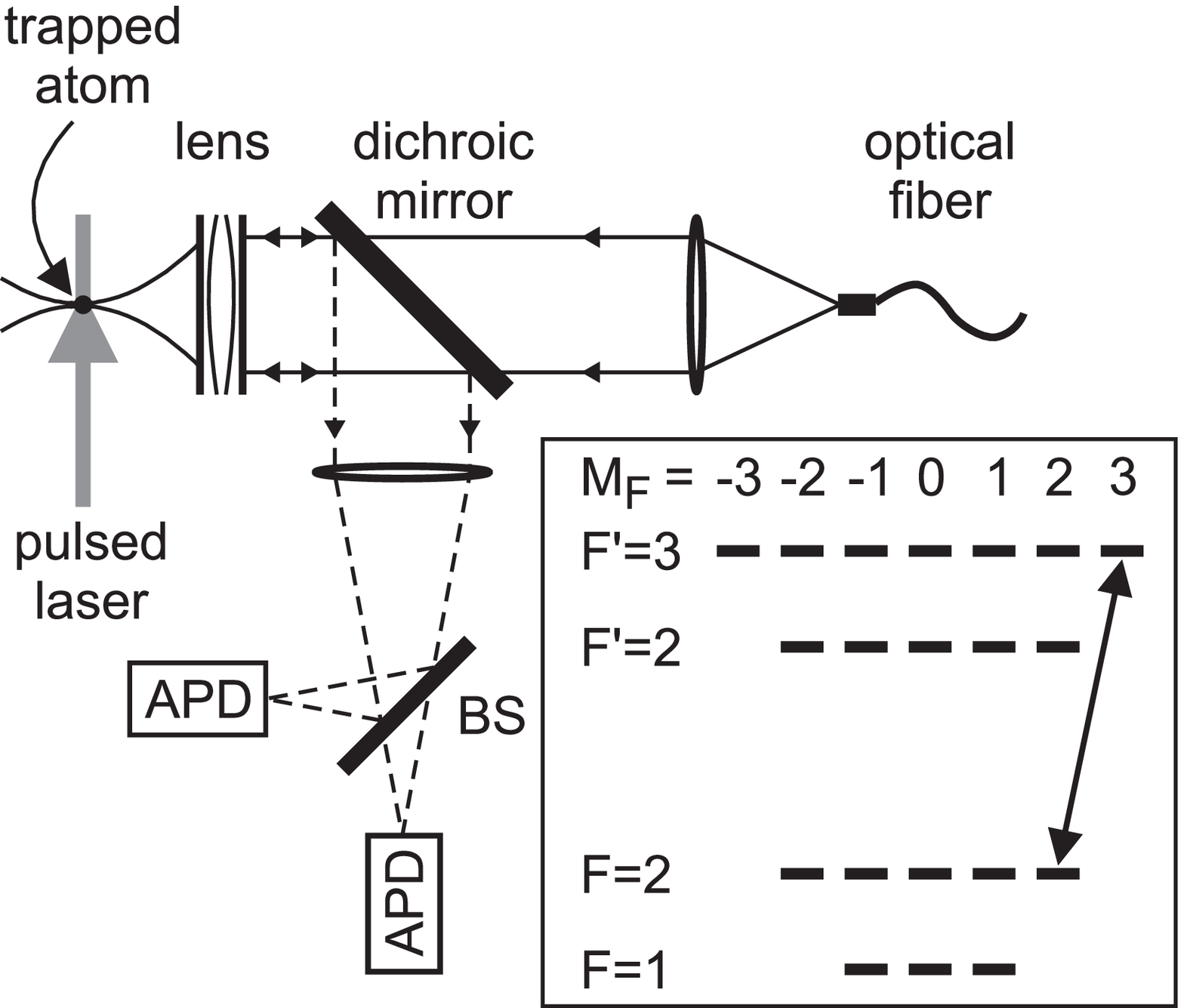}
\end{figure}
\begin{figure}
\includegraphics{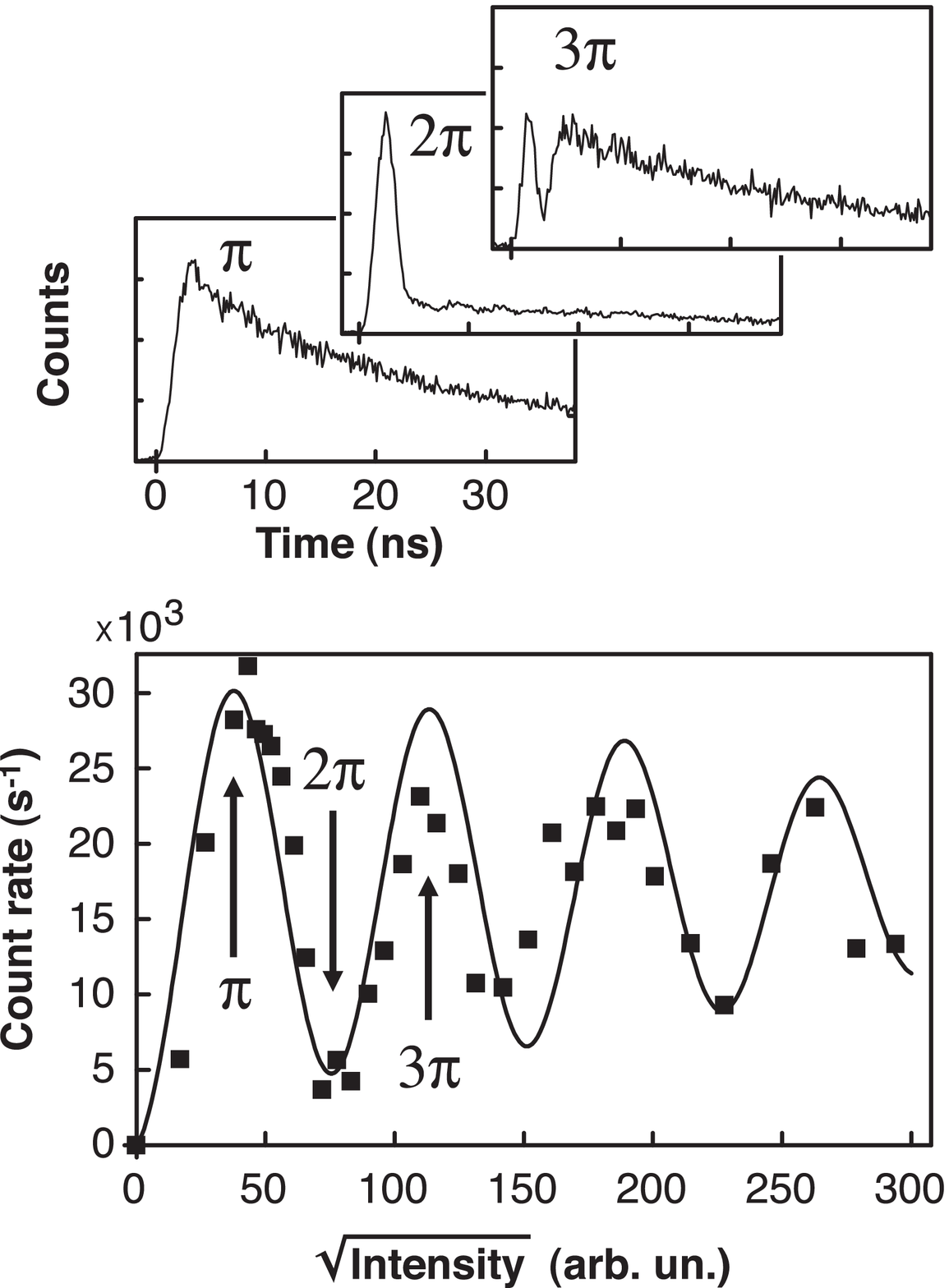}
\end{figure}
\begin{figure}
\includegraphics{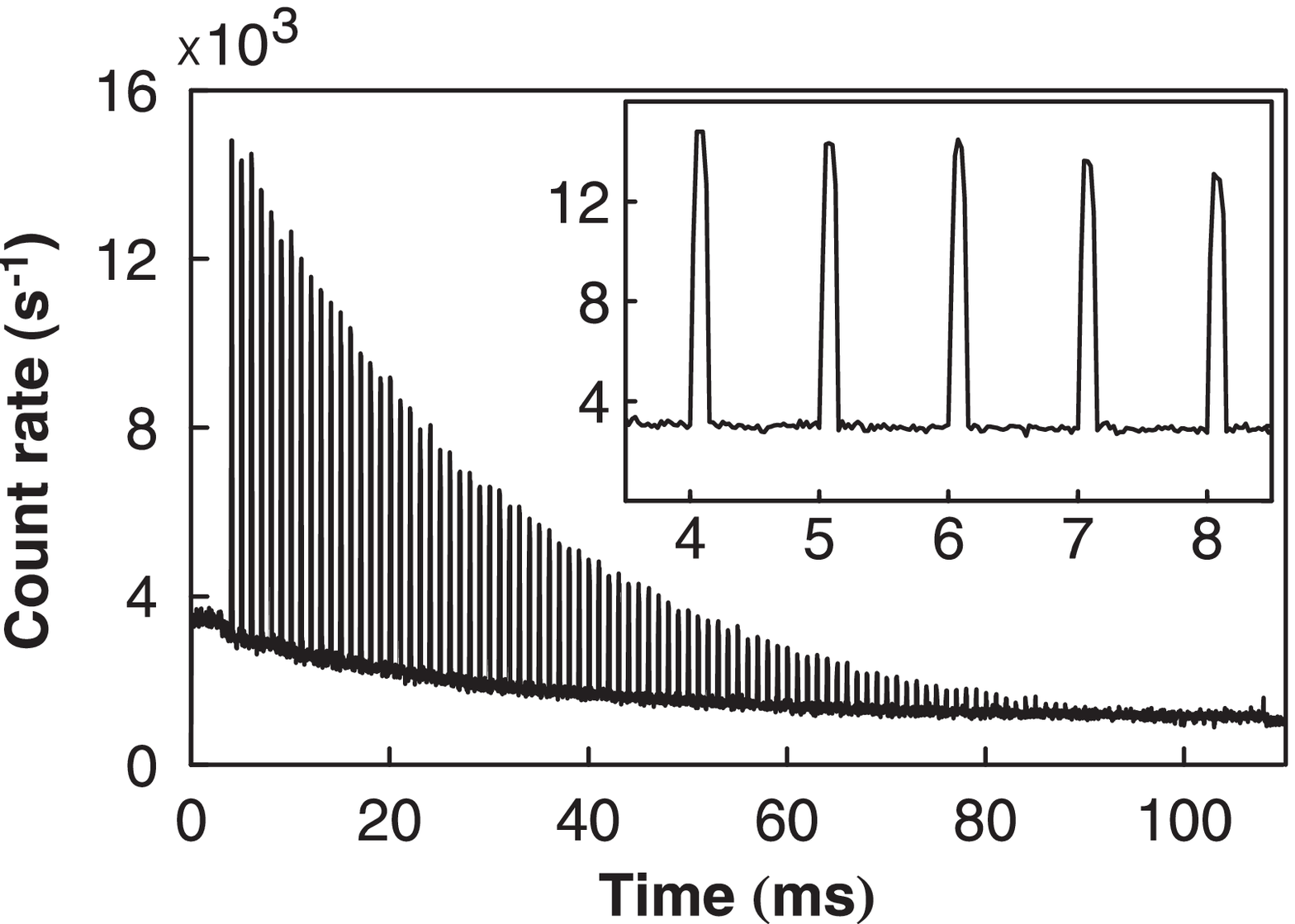}
\end{figure}
\begin{figure}
\includegraphics{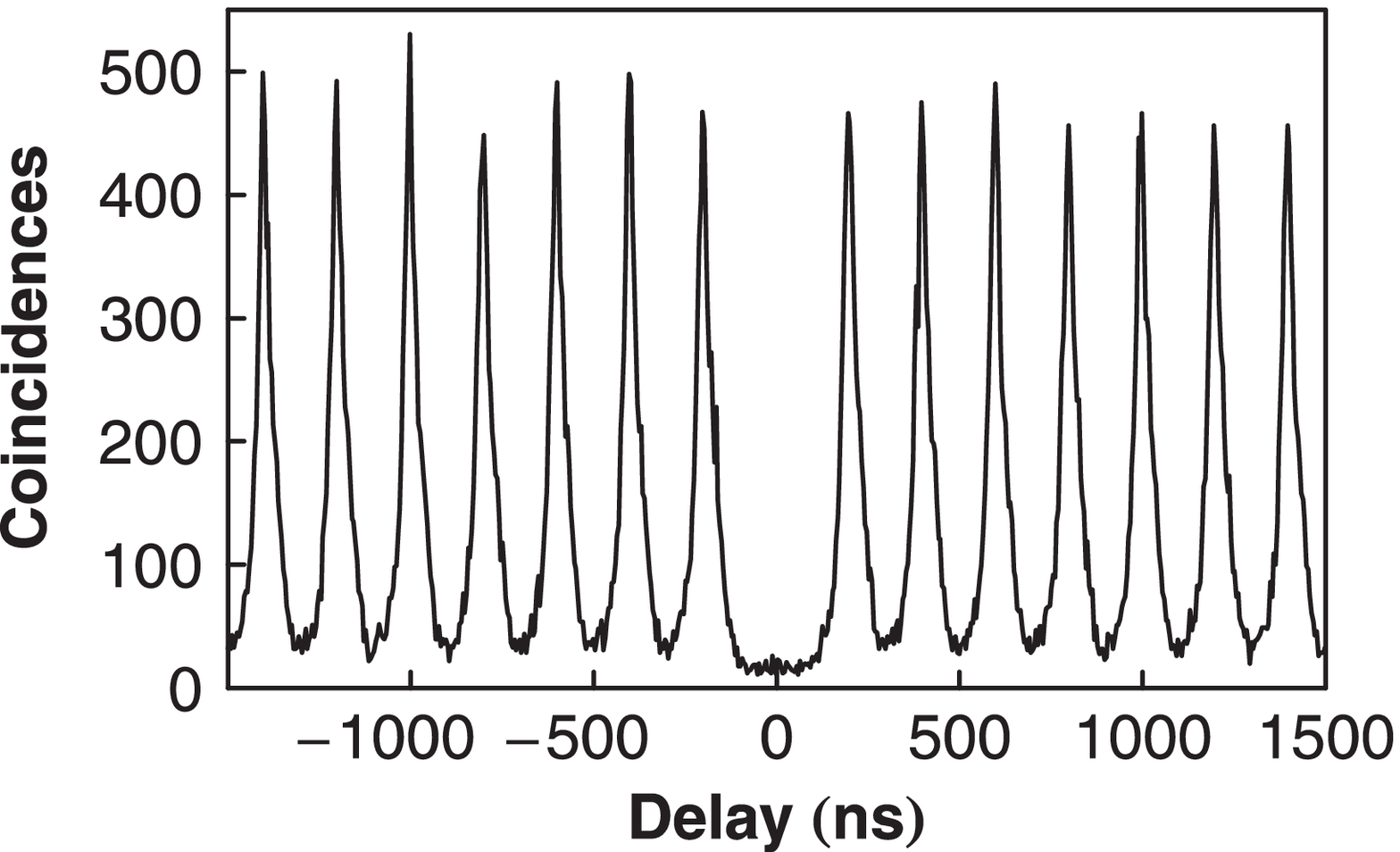}
\end{figure}
\end{document}